\documentclass[aps,twocolumn,10pt,nofootinbib]{revtex4}
\usepackage{graphicx,amssymb,amsmath,subfigure}
\newcommand{\ket}[1]{\left|#1\right>}

\newcommand{\braket}[1]{\left<#1\right>}
\newcommand{\para}[1]{\left(#1\right)}
\newcommand{\abs}[1]{\left|#1\right|}

\newcommand{\sd}[0]{\ \ \ }

\begin{document} 
\title{Fractional topological superconductors with fractionalized Majorana fermions}
\author{Abolhassan Vaezi}
\affiliation{School of Physics, Institute for research in Fundamental Sciences (IPM), 19395-5531, Tehran, IRAN\\
and\\
Department of Physics, Cornell University, Ithaca, New York 14853, USA}
\email{vaezi@cornell.edu}


\begin{abstract}
In this paper, we introduce a two-dimensional fractional topological superconductor (FTSC) as a strongly correlated topological state which can be achieved by inducing superconductivity into an Abelian fractional quantum Hall state, through the proximity effect. When the proximity coupling is weak, the FTSC has the same topological order as its parent state and is thus Abelian. However, upon increasing the proximity coupling, the bulk gap of such an Abelian FTSC closes and reopens resulting in a new topological order: a non-Abelian FTSC. Using several arguments we will conjecture that the conformal field theory (CFT) that describes the edge state of the non-Abelian FTSC is $U(1)/Z_2$ orbifold theory and use this to write down the ground-state wave function. Further, we predict FTSC based on the Laughlin state at $\nu=1/m$ filling to host fractionalized Majorana zero modes bound to superconducting vortices. These zero modes are non-Abelian quasiparticles which is evident in their quantum dimension of $d_m=\sqrt{2m}$. Using the multi-quasi-particle wave function based on the edge CFT, we derive the projective braid matrix for the zero modes. Finally, the connection between the non-Abelian FTSCs and the $Z_{2m}$ rotor model with a similar topological order is illustrated.
\end{abstract}

\maketitle

\section{Introduction}
There is growing interest in strongly correlated topological states of matter and their potential application in topological quantum computation.\cite{Rev_1, Das_Sarma_1,Rezayi_2010_1,PC_0,FTI_1,ITI_1,ITI_2,ITI_3,S_1,S_2,Fendley_2012_1} Interaction can lead to novel topological states exhibiting fractional excitations, anyon statistics of quasiparticles, and nontrivial ground-state degeneracy (GSD).\cite{Wen_book_1,Wen_1990_1} Interestingly, these properties of the topological states are robust and insensitive to disorder or interaction as long as the bulk gap does not close.\cite{Rev_1} An appealing topological state of matter is the non-Abelian phase that supports anyon excitations with quantum dimensions more than one that obey non-Abelian braid statistics.\cite{Read_91,Wen_1991_1,Read_Rezayi_1999} This exciting possibility along with the robustness of the topological states that immunizes anyons from quantum decoherence make non-Abelian anyons a promising candidate for topological quantum computation \cite{Rev_1}. The first non-Abelian topological phase was proposed  by Read and Moore for the fractional quantum Hall (FQH) state at $\nu=1/2$ filling fraction. Their so-called Pfaffian wave vfunction is a paired state of composite fermions whose excitations obey Ising-type non-Abelian statistics. \cite{Read_91} It was shown later that the projective Ising-type non-Abelian statistics can also be realized in the $p+ip$ superconductors in the weak pairing phase. \cite{Read_2000,Ivanov} Indeed, the Pfaffian FQH state and a $p+ip$ superconductor have the same topological order, whose excitations are the celebrated Majorana fermions with quantum dimensions equal to $\sqrt{2}$.

\begin{figure}[tbp]
\begin{center}
\includegraphics[width=250pt]{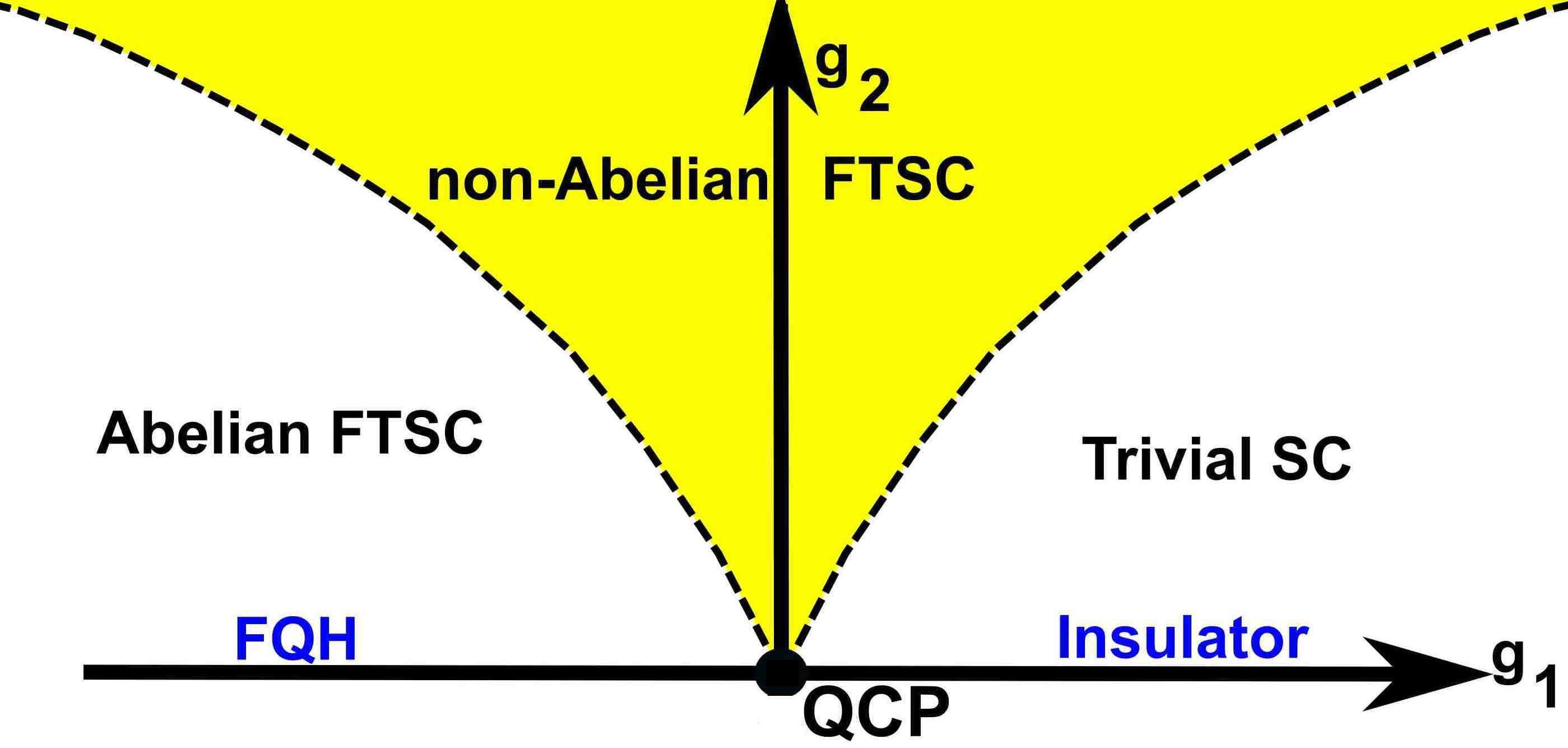}
\caption{A schematic phase diagram of the FTSCs. $g_{1}$ is a tuning parameter that drives the phase transition between an FQH at $\nu=1/m$ filling fraction and a trivial insulating state, e.g., a Wigner crystal. $g_2$ denotes the tunneling amplitude between the superconducting substrate and the FQH sample. The edge theory of the Abelian FTSC side is described by a free boson compactified on a circle with radius $R=\sqrt{m}$, i.e., $U(1)_{m}$ CFT, while the edge state of the quantum critical region i.e. the non-Abelian side, is given by the $U(1)_{m}/Z_2$ orbifold CFT. The trivial superconductor does not support gapless edge states.}\label{Fig02}
\end{center}
\end{figure}

Recently, it has been shown that the $p+ip$ superconductor can be achieved by inducing superconductivity in the bulk of an integer  quantum anomalous Hall (QAH) state via proximity to a superconductor.\cite{Fu_2008_1,Qi_2010_0} In a different study, Bombin showed that Majorana fermions can emerge in an Abelian topological state with topological defects. \cite{ Bombin_2010} These two mechanisms can be related through viewing superconducting vortices as twist operators. These observations motivated us to investigate the fate of an Abelian FQH state or a fractional Chern insulator at $\nu=1/m$ filling fraction in a similar setup. It will be shown that the weak tunneling of Cooper pairs does not affect the topological nature of the FQH state. We refer to the resulting phase as the Abelian fractional topological superconductor (FTSC). However, we discuss that the system can undergo a  topological phase transition by increasing the tunneling strength. The new topological state will be a non-Abelian FTSC, whose excitations are fractionalized Majorana fermion zero modes, $\alpha_0$. These non-Abelian excitations are bound to the superconducting vortices,  follow $\alpha_{0}^{2m}=1$ relation, and have quantum dimension $d_m=\sqrt{2m}$.

The paper is organized as follows. In Sec. II, we define the notion of a FTSC and try to build a model for it. To make the situation more comprehensible, an exactly solvable model will be reviewed in which a topological phase transition happens between an integer QAH and a $p+ip$ superconductor. In Sec. III, we identify the CFT that describes the edge theory of the FTSCs. The correspondence between the bulk wave function and the conformal blocks of the edge theory is explained, and the notion of the twist field is illustrated with an emphasis on their crucial role in the topological order. In Sec. IV, we utilize the CFT that describes the non-Abelian FTSCs to study its topological order. The existence of the zero modes bound to vortices of the superconducting state is proven. Furthermore, the quantum dimension and the braid statistics of the zero modes is derived. Sec. V is devoted to the computation of the bulk wave function of the non-Abelian FTSCs. In Sec. VI, we study the gauge theory of the non-Abelian FTSCs. It turns out to be a $Z_{2m}$ gauge theory which allows us to make a connection between our prediction about the quantum dimension of the fractionalized Majorana zero modes and the results of the recently studied rotor model with a similar gauge structure. 

\section{What are fractional topological superconductors?}

In this section, we try to present a definition of a FTSC and what exactly we mean by this term. Roughly speaking, an FTSC is a superconducting state of matter whose parent state is an FQH state. Two topologically distinct classes of FTSCs can be imagined, Abelian and non-Abelian FTSCs. Abelian FTSCs are the counterpart of the Abelian FQH states and refer to the case where all the excitations of the superconducting ground state are Abelian anyons. In other words, exchanging any of two quasiparticles can only change the phase of the ground-state wave function, and two successive exchanges are commuting. Accordingly, all quasiparticles have quantum dimensions equal to one. An important signature of the Abelian FTSCs is the uniqueness of their ground-state wave functions in the presence of vortices. On the other hand, non-Abelian FTSCs have non-Abelian excitations with non-Abelian statistics. The quantum dimension of a non-Abelian quasiparticle (non-Abelion) is necessarily more than one leading to degenerate ground states. Exchanging two non-Abelions not only can change the phase of the ground state, it may also act on the space of ground states by a unitary transformation. In other words, the ground state we started with can superpose with other degenerate ground states after exchanging two non-Abelions.  

Based on what we said in the above paragraph, an Abelian FTSC has the same topological order as an Abelian FQH state. In other words, if we break all the symmetries of both phases, they are smoothly connected; i.e. one of them can evolve to the other one adiabatically without passing any phase transition. An FTSC has all the symmetries of its parent FQH state , except the particle number conservation [U(1) symmetry]. The U(1) symmetry can be easily broken by inducing superconductivity in the FQH state either intrinsically or extrinsically. In the latter scenario, the FQH system can be put above a superconductor, so that Cooper pairs can tunnel into the FQH state through proximity effect.  If  we keep the many-body gap open in the whole procedure, the resulting state will be an Abelian FTSC. Consequently, we expect the same kind of chiral edge theory for the system as that of the parent FQH, except that Abelian anyons do not carry a definite electric charge anymore. A more interesting situation can happen when the many-body gap closes due to the proximity and the system undergoes a topological phase transition accordingly. The important question that we are going to address in this paper is the fate of the Abelian state. In the absence of the superconductivity, the fate of an FQH is more or less known.  For instance, upon decreasing the filling fraction, e.g., through tuning the chemical potential, we can end up the Wigner crystal phase at small filling fractions. Now, let us consider the quantum phase transition between the FQH state and the trivial insulating phase which can be a Wigner crystal, for example. A weak proximity coupling turns the two sides of the phase transition an Abelian FTSC and a trivial superconductor with no edge state, respectively. However, the strong tunneling of Cooper pairs can open a quantum critical region between these two regions. As is shown later, we conjecture through different arguments and analogies that the intervening state is an FTSC with non-Abelian excitation. It is shown that when inducing superconductivity in the Laughlin state at $\nu=1/m$ filling, the vortices host non-Abelian fractionalized Majorana zero modes with quantum dimension $d_{m}=\sqrt{2m}$.

\subsection{Superconductivity proximity effect on integer quantum Hall states}

In this section, we discuss the superconductivity proximity effect  in a chiral model of the noninteracting electrons whose phase diagram resembles the physics we are pursuing. This simple model has been proposed by Qi {\em et al}.\cite{Qi_2010_0} A similar model has also been studied by Fu and Kane.\cite{Fu_2008_1}  Before writing the Hamiltonian, we would like to point out some comments that makes the results more comprehensible. Imagine an integer QAH system, e.g., a band structure with Chern number $C=1$.  The edge theory of this system can be described by chiral electrons (or free bosons) propagating along the edges of the system and localized in the normal direction.  By tuning the mass term (in this model the magnetization of electrons), we can undergo a phase transition into a trivial state with $C=0$. At the critical point, the system becomes gapless and therefore, an electron at the upper edge, which is left mover, can hybridize with (or tunnel into) the right mover lower edge state. This can happen since the bulk state is gapless at the critical point and edge electrons can hybridize with the bulk electrons to travel from one side to the other. This coupling between the upper and lower edges can destabilize the edge state and gap it out. When gap reopens by increasing chemical potential, the system becomes a trivial phase with no gapless edge state. 

Now, let us induce superconductivity in the whole system, {\em i.e. in the bulk as well as the edge}. In the weak coupling regime where only a few Cooper pairs can tunnel into the system and cannot affect the band-gap, the situation is clear. It turns the  QAH state into an Abelian topological superconductor whose edge state is still described by chiral {\em complex fermions} (Bogoliubov quasiparticles). Complex fermions, e.g. electrons have two degree of freedom, the real and the imaginary part of fermion operator.  Each of these degrees of freedom can be represented by a Majorana operator, which is a {\em real fermion}. By computing the Chern number of the resulting Abelian superconductor, we would obtain $C=2$ referring to the fact that the edge state can be described by two Majorana fermions or equivalently a single complex fermion, in this case Bogoliubov quasiparticles. Moreover, inducing superconductivity into the trivial phase via proximity effect simply turns it to a trivial superconductor. Now let us discuss what happens in the strong coupling between the superconductor and the QAH system. By strong we mean that the induced Cooper pairs can close and reopen up the band-gap in a new fashion. The best way to think about this problem is to consider the edge theory when the bulk gap closes. Like before, left-mover edge electrons at the upper boundary can tunnel into the lower edge state with right-mover electrons through the gapless bulk and hybridization with bulk electrons  that obviously causes instability. However, note that we have a different instability channel, the Cooper pairing, as well. {\rm} The upper left mover edge electrons can pair with the right mover edge electron, since the gapless bulk is a superconducting state and can mediate Cooper pairs. It is this Cooper pairing channel that provides novel possibilities. Now let us look more carefully. Approaching from the Abelian topological superconducting region, the edge theory is described by a single complex fermion or equivalently two real (Majorana) fermions. Through the pairing between the upper and lower edges, we can gap out one of the real (Majorana) fermions and keep the other gapless. Therefore, we end up with a topological superconductor whose edge theory is given by a {\em single Majorana fermion mode}. Since Majorana fermions obey non-Abelian statistics and have quantum dimension $d=\sqrt{2}$, the intervening state is a non-Abelian integer superconductor. Now, let us present the model Hamiltonian which is given by the following equation:

\begin{eqnarray}
  &&H_{QAH}=\sum_{p}=\psi_p^\dag h_{QAH}\para{p}\psi_p ,\cr
  &&h_{QAH}\para{p}=\left(
                      \begin{array}{cc}
                        m\para{p} & A\para{p_x-ip_y} \\
                        A\para{p_x+ip_y} & -m\para{p} \\
                      \end{array}
                     \right),
\end{eqnarray}
where $m\para{p}=m_0+Bp^2$ and $\psi_p=\para{c_{p,\uparrow},c_{p,\downarrow}}^{\rm T}$. This type of coupling between spin-up and spin-down electrons can be achieved, for example, in the Rashba-type spin-orbit coupled system, e.g., at the surface of the three-dimensional topological insulators. $m_0$ represents the average magnetization of electrons that can be realized either through the Zeeman-type interaction with an external magnetic field or the exchange interaction with a ferromagnet on the top.  Assuming all the negative energy states are occupied and $\abs{m_0}> 0$, the Hall conductivity of the above Hamiltonian is given by $\sigma_{H}=C\frac{e^2}{h}$, where $C$ is the Chern number of the lower energy band. It can be shown that $C=+1$ when $m_0<0$ and vanishes for $m_0>0$. Now we turn the above QAH system to a TSC. An effective pairing interaction between quasiparticles can be induced in the proximity of an $s$-wave superconductor. The corresponding BogoliubovÐde Gennes (BdG) Hamiltonian is

\begin{eqnarray}
  H_{BdG}=\frac{1}{2}\sum_{p}\Psi_p^\dag \left(
                                           \begin{array}{cc}
                                             h_{QAH}\para{p}-\mu & i\Delta \sigma^y \\
                                             -i\Delta^*\sigma^y & -h^*_{QAH}\para{-p}+\mu \\
                                           \end{array}
                                         \right)
  \Psi_p\cr
  &&
\end{eqnarray}
where $\Psi_p=\para{c_{p,\uparrow},c_{p,\downarrow},c_{-p,\uparrow}^\dag,c_{-p,\downarrow}^\dag}^{\rm T}$ and the chemical potential $\mu$ has been introduced so that we can dope the unpaired QAH state. Intuitively, we expect that the ground state becomes a topological superconductor for $\abs{\Delta}>\abs{m_0}$ and $\mu=0$. The full phase diagram of the above Hamiltonian has been derived in Ref. \cite{Qi_2010_0}. There are several interesting facts about them, which we mention below.

\subsection*{$\mu=0$ case}

(A) For $\Delta=0$, there is a phase transition at $m_0=0$. For $m_0<0$, the edge state hosts a chiral Luttinger liquid. This edge state which has central charge $c=1$, can be written in terms of two chiral Majorana modes each having $c=\frac{1}{2}$. However, for $m_0>0$ there is no edge state. Therefore, both Majorana edge states gap out through the transition at $\mu=\Delta=0$.

(B) For nonzero values of $\Delta$, when $\Delta<-m$, we still have two gapless Majorana modes at the edge. The wave function of these Majorana modes is extended along the edge and localized in the perpendicular direction. As $\Delta$ approaches $-m$ from below, the localization length of one Majorana mode normal to the edge becomes larger and diverges at $\Delta=-m_0$. However, the other Majorana mode is still localized normal to the edge, and extended along it. Hence, at $\Delta=-m_0$, only one of the Majorana modes delocalizes and merges with the Bulk states. When $-m_0<\Delta<m_0$, that Majorana mode becomes gapped and only the other zero mode survives. As we approach $\Delta=m_0$, the localization length for the remaining Majorana mode normal to the edge increases, and finally diverges at $\Delta=m_0$.  For $m_0>\Delta$, the only remaining Majorana modes gaps out as well. {\em Therefore, a topological superconductor emerges at the critical quantum point between a QAH state and a trivial insulator}.

(C) The above transition can be described in the following way. In the absence of pairing potential ($\Delta=0$), both Majorana modes have the same correlation length perpendicular to the edge (we estimate the width of a Majorana mode by this length scale) which is $\xi \propto \frac{1}{\abs{m_0}}$. As a result, both will be gapped at the same critical point. However, turning on a nonzero $\Delta$, the correlation lengths of two Majorana modes differ and we have $\xi_1\propto 1/\abs{m_0+\Delta}$ and $\xi_2 \propto 1/\abs{m_0-\Delta}$. At $\Delta=- m_0$, the first one has an infinite correlation length and delocalizes, while the other one has a finite correlation length. The first Majorana zero mode merges with bulk states and disappears beyond that point. At $\Delta=+ m_0$, the remaining Majorana mode acquires infinite correlation length and gaps out accordingly.

(D) A similar phase diagram is expected for the proximity of an integer quantum Hall (QH) systems (filled Landau levels) to an $s$-wave superconductor. The only topological index that labels the integer QAH or QH states is their Chern number; therefore, they are topologically indistinguishable. The same edge theory describes both, so we expect the same topological phase transitions in both cases.  

\subsection*{$\abs{\mu} > 0$ case}

In Ref. \cite{Qi_2010_0}, a phase diagram has been presented for this case. However, we argue that their result is valid only in the weakly correlated systems where either band structure is far from being flat or interaction is weak. Going back to the result of Qi {\em et al}, at $\Delta=0$, nonzero chemical potential yields a metallic state. However, turning on the $\Delta$ term, chemical potential changes only the phase transition point given by the following expression:

\begin{eqnarray}
&& \Delta^2+\mu^2=m^2.
\end{eqnarray}

\begin{figure}[tbp]
\begin{center}
\includegraphics[width=200pt]{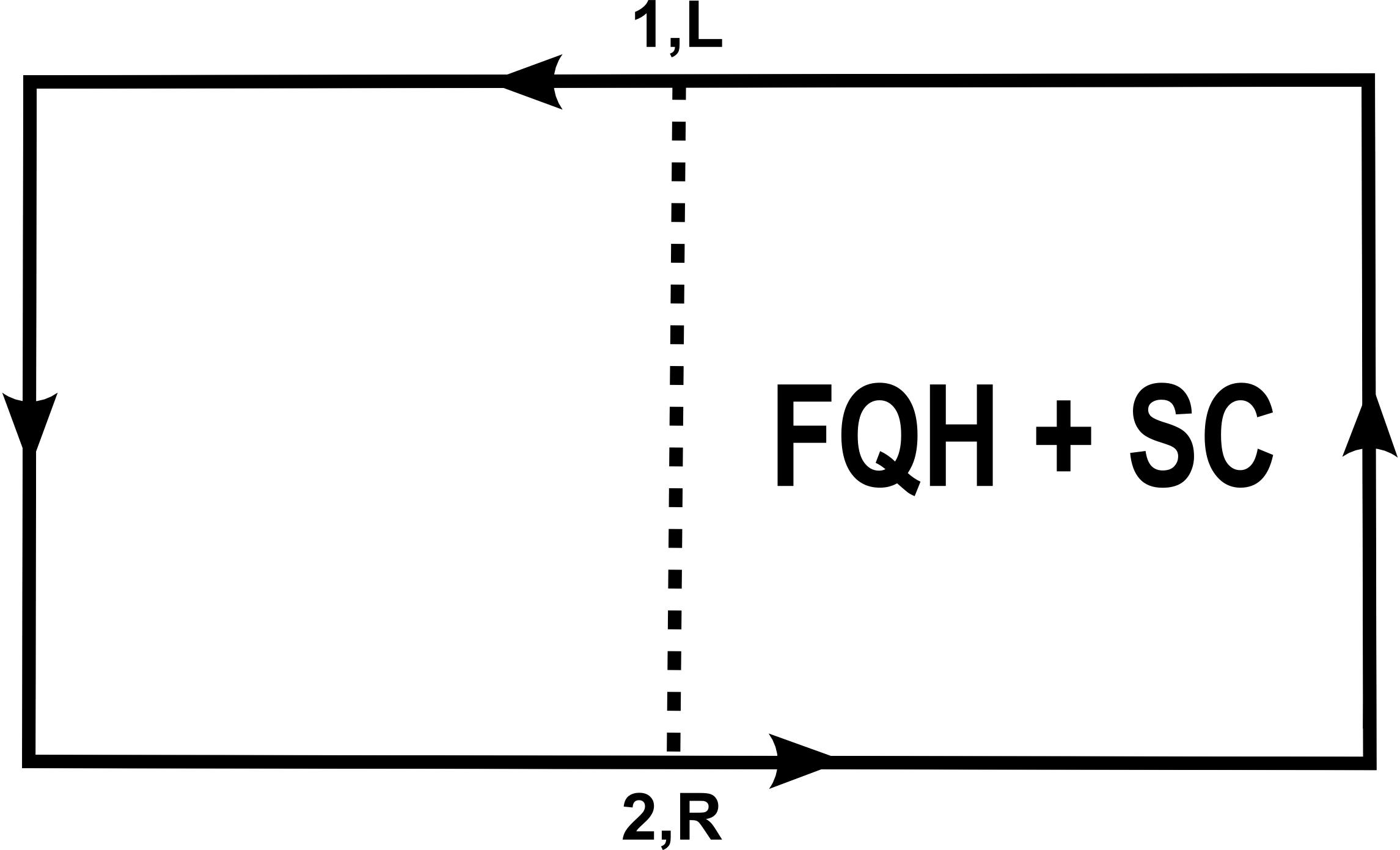}
\caption{The quantum critical point (QCP) between an Abelian and a non-Abelian FTSCs whose parents are $1/m$ Laughlin state. At the QCP, the bulk gap vanishes and the localization length of the edge state diverges. As a result, the top ($1,L$) and bottom ($2,R$) sides of the edge state that are counterpropagating can interact in the Cooper channel represented by the dashed line. Approaching the QCP from the Abelian side, the edge quasiparticles are complex anyons  with two degrees of freedom: $V_l= e^{il\varphi/\sqrt{m}}$, where $l\in Z_{m}$. This pairing between ($1,L$) and ($2,R$)  can gap out the $\para{V_l-V_l^\dag}/2i$ part of the anyon field. Accordingly, only the remaining part, $\para{V_l+V_l^\dag}/2=\cos\para{l\varphi/ \sqrt{m} }$, which is invariant under the $Z_2$ action, $\varphi \to -\varphi$, remains gapless. This argument suggests that the edge theory of the resulting state after gap reopens is achieved by taking $\cos\para{\sqrt{m}\varphi}$ as the electron operator instead of $e^{i\sqrt{m}\varphi}$. The CFT that contains this electron operator, and with central charge $c=1$, is known to be a $U(1)_{m}/Z_2$ orbifold theory.\cite{Vafa_1,PC_1} Interestingly, such an orbifold CFT contains non-Abelian primary fields with quantum dimensions $\sqrt{2m}$. } \label{Fig01}
\end{center}
\end{figure}

\subsection{Generalization to the fractional case}

Let us look more carefully at the doped QAH state in the proximity of an $s$-wave superconductor, discussed above. Assuming a nearly flat lowest band for the Hamiltonian describing the QAH state $H_{QAH}$ and doping this system so that the filling fraction becomes fractional, e.g., $\nu=\frac{1}{m}$, we can add an interaction whose coupling constant is significantly larger than the bandwidth and significantly smaller than the energy gap between lower and upper bands. This is the definition of a nearly flat Chern insulator. Although, doping leads to a metallic behavior in the integer Chern insulators, in the present case the many body wave function becomes nontrivial and exhibits topological order. The excitations carry fractional charge $e^*=\nu$ at $\nu=\frac{1}{m}$ filling fraction, and they have fractional statistics with $\theta=\frac{\pi}{m}$. Finally, the GSD on the torus is $m$. \cite{CI_1,CI_2,CI_3,CI_4,Vaezi_2011_05,Qi_2011_1} When the system has an open boundary, the edge state is described by a chiral boson mode $\varphi\para{x,t}$, from which we can construct the electron operator $\Psi\para{x,t}=e^{i\sqrt{m}\varphi\para{x,t}}$. Similarly, the anyon operator with electric charge $\pm e\frac{l}{m}$ can be represented in terms of the free chiral boson as $V_{\pm l}=e^{\pm i\frac{l}{\sqrt{m}}\varphi\para{x,t}}$. These complex anyon operators have two degrees of freedom, $\cos{\frac{l}{\sqrt{m}}\varphi\para{x,t}}$ and $\sin{\frac{l}{\sqrt{m}}\varphi\para{x,t}}$, which are eigenstates of the following $Z_2$ symmetry operation: $\mathcal{R}: \phi\to -\phi$. Interestingly, these primary fields do not have a definite electric charge and are suitable when the electromagnetic gauge symmetry is broken, i.e. in a superconductor. The key assumption which is used extensively in the rest of this paper is that {\em above a critical pairing strength, only one of these $Z_2$ eigenstates survives} and the other becomes gapped. Therefore, the electron operator at the edge is $V_{e}=\frac{\Psi+\Psi^\dag}{2}=\cos\para{\sqrt{m}\varphi}$, which is self-conjugate, i.e., $V_{e}\times V_{e}\sim 1$. The CFT that contains this electron operator at $c=1$ has been studied extensively in the CFT literature and is called $U(1)_{m}/Z_2$ orbifold theory. \cite{Vafa_1,Ginsparg_1991,DiFrancesco_1,PC_0,PC_1,Barkeshli_2012_1} This orbifold is known to have non-Abelian primary fields, suggesting the name non-Abelian FTSC for the bulk state. It is worth mentioning that the non-Abelian FTSC can also be obtained by inducing superconductivity in the FQH states, and the procedure explained above is not specific to fractional QAH systems.

\section{CFT approach to the ${\rm \bf FTSCs}$}

\subsection{Bulk wave function and correlation functions of the edge CFT}
For the chiral topological states, it is conjectured that the holomorphic part of the bulk wave function can be described as the correlation function of the fermion operators in the chiral edge theory.\cite{Read_91,Read_Rezayi_1999,Shankar_Ashvin_2011} First of all, we have to identify the CFT that describes the chiral edge state. Second, we should find the electron operator $V_{e}\para{z}$ that has a half integer conformal dimension. This is to makes sure that two electron operators anticommute. Then the wave function of the ground state of $N$ electrons with electrons coordinates $\{z_k\}$, where $z_k=x_k+iy_k$,  is simply given in terms of the correlation function involving $N$ electrons operators at $\{z_k\}$ so $\Psi\para{\{z_k\}}=\braket{\prod_{k=1}^{N}V_{e}\para{z_k}}$. Third, we can construct the excited wave function by inserting other primary fields in the correlation wave function. For example, in the $1/m$ Laughlin states, the electron operator is $V_e=e^{i\sqrt{m}\varphi}$ and  $V_{a}=e^{i\frac{\varphi}{\sqrt{m}}}$ is the primary field associated with the fractional excitation whose charge is $q=e/m$. Hence, the holomorphic part of the wave function for $N$ electrons and $n$ anyons would be of the following form:
\begin{eqnarray}
&&\Psi\para{\{\eta_l\};\{z_k\}}=\braket{\prod_{l=1}^{n}V_{a}\para{\eta_{l}}\prod_{k=1}^{N}V_{e}\para{z_k}}=\cr
&&\prod_{i,l}\para{z_i-\eta_l}\prod_{k<l}\para{\eta_l-\eta_k}^{1/m}\prod_{i<j}\para{z_i-z_j}^{m}.
\end{eqnarray}
In the next section, we present a comprehensive calculation of the steps explained above for the non-Abelian FTSCs and identify the primary fields of the edge CFT, the fusion algebra of that CFT, and the bulk wave-function. We also study the non-Abelian statistics of the fractionalized Majorana zero modes and compute the corresponding braid matrix.
\subsection{Edge theory of the Abelian FTSCs}

As we mentioned before, an Abelian FTSC is smoothly connected to its parent $FQH$ state at the same filling fraction. Thus, they belong to the same topological class and have identical edge theories, except that the number conservation is broken at the edge of an FTSC. So it is enough to consider the edge CFT of an FQH state. The edge of a Laughlin state at $\nu=1/m$ is described by a chiral gapless mode given by the following action\cite{Wen_book_1}:

\begin{eqnarray}
  S=-\frac{m}{4\pi}\int dxdt~ \para{\partial_t+v_{\rm F}\partial_x}\varphi\para{x,t}\partial_{x}\varphi\para{x,t}.
\end{eqnarray}
The above edge theory can be written in terms of a $U(1)$ CFT on the 2D space-time. To this end, we use the following notations:

\begin{eqnarray}
z=t+ix\sd,\sd \bar{z}=t-ix\sd,\sd x\sim x+2\pi
\end{eqnarray}
where the periodicity of the $x$ direction is due to the circular geometry of the edge~($x$ is rescaled to satisfy the above relation and the Fermi velocity is chosen as $v_{\rm F}=1$). By this transformation, the action of the edge theory becomes 
\begin{eqnarray}
  S=-\frac{1}{4\pi}\int dzd\bar{z}~ \partial_{z} \varphi\para{z,\bar{z}} \partial_{\bar{z}} \varphi\para{z,\bar{z}},
\end{eqnarray}
in which we have rescaled $\varphi$ by a factor of $\sqrt{m}$. We should be careful about this scaling as it has a physical effect. Before its action, the bosonic $\varphi$ is assumed to be living on a circle with radius $R=1$;  i.e., $\varphi $ is identified with $\varphi+2\pi$. However, after it we should take $R=\sqrt{m}$, i.e., $\varphi \sim \varphi+2\pi \sqrt{m}$. Technically speaking the bosonic field is compactified on a circle with radius $R=\sqrt{m}$.

The equation of the motion for the above action is $\partial_{z} \partial_{\bar{z}} \varphi\para{z,\bar{z}}=0$, so either $\partial_{z} \varphi\para{z,\bar{z}}=0$ or $ \partial_{\bar{z}} \varphi\para{z,\bar{z}}=0$. We choose the latter choice, which means $\varphi\para{z,\bar{z}}=\varphi\para{z}$ since we need a chiral (holomorphic) edge state (the first choice given an antichiral (nonholomorphic) edge state. The edge state is enjoying a $U(1)$ symmetry under which $\varphi\to \varphi+\epsilon$ with any arbitrary  $\epsilon$.  This symmetry leads to a conserved Noether current. It is given by the following formula:

\begin{eqnarray}
j\para{z}=i\partial_{z}\varphi\para{z},
\end{eqnarray}
which is manifestly conserved since $\partial_{\bar{z}} j\para{z}=i\partial_{\bar{z}}\partial_{z}\varphi\para{z}=0$.  The propagator of the $\varphi$ field satisfies the following equation:

\begin{eqnarray}
&&-\frac{1}{2\pi}\partial_{\bar{z}}\partial_{z}K\para{z,\bar{z},w,\bar{w}}=\delta^{\para{2}}\para{z-w}.
\end{eqnarray}
Using the $\frac{1}{2\pi}\partial_{z}\para{\bar{z}-\bar{w}}^{-1}$ identity for the delta function in complex plane, we have:
\begin{eqnarray}
K\para{z,\bar{z},w,\bar{w}}&&=\braket{\varphi\para{z,\bar{z}}\varphi\para{w,\bar{w}}}=-\ln \abs{z-w}^2\cr
&&=-\ln \para{z-w}-\ln \para{\bar{z}-\bar{w}}.
\end{eqnarray}
For holomorphic solutions of the  edge theory we have $\braket{\varphi\para{z}\varphi\para{w}}=-\ln \para{z-w}$. Now let us find the correlation function of the current operators.
\begin{eqnarray}
&&\braket{j\para{z} j\para{w}}=-\partial_{z}\partial_{w}\braket{\varphi\para{z}\varphi\para{w}}=\frac{1}{\para{z-w}^2}.
\end{eqnarray}
We can read the scaling (conformal) dimension of current operator off above correlation means leading to $h_{j}=1$.  An interesting set of operators known as vertex operators can be obtained from the bosonic $\varphi$ chiral fields. We have

\begin{eqnarray}
&&V_{\alpha}\para{z}=e^{i\alpha\phi\para{z}}.
\end{eqnarray}
It can be easily shown that
\begin{eqnarray}
&&\braket{V_{\alpha}\para{z}V_{\beta}\para{w}}=\para{z-w}^{\alpha\beta}.
\end{eqnarray}
An important fact about these operators is that they satisfy $\braket{V_{\alpha}\para{z}V_{\beta}\para{w}}=e^{i\pi\alpha\beta}\braket{V_{\beta}\para{z}V_{\alpha}\para{w}}$, which gives anyonic statistics for $\alpha\beta \notin {Z}$. Taking $\alpha=\beta$ we can compute the scaling dimension of $V_{\alpha}$, which happens to be

\begin{eqnarray}
&&h_{\alpha}=\alpha^2/2.
\end{eqnarray}
The fact that $\varphi$ is compactified on a circle with $R=\sqrt{m}$ allows only discrete values for $\alpha$. We have

\begin{eqnarray}
V_{\alpha}\para{z}&&=e^{i\alpha\phi\para{z}}=e^{i\alpha\para{\phi\para{z}+2\pi \sqrt{m}}} ~\Rightarrow \cr
&&\alpha=\frac{k}{\sqrt{m}}~,~ k\in {Z}.
\end{eqnarray}
Note that there is no constraint on the value of the $k$ and it can be any integer number. Therefore, we still have an infinite number of primary fields. In the next section, we discuss the chiral algebra that puts further constraints on the value of $k$. Now let us consider the following operator:

\begin{eqnarray}
V_{e}\para{z}=e^{i\sqrt{m}\varphi\para{z}}\sd, \sd V_{e}^\dag\para{z}=e^{-i\sqrt{m}\varphi\para{z}}.
\end{eqnarray}
The scaling dimension of this operator is $m/2$ and exchanging two of them yields a minus sign. Accordingly, $V_{e}\para{z}$ represents a fermion operator. By looking at the operator product expansion (OPE) between $V_{e}$ and the current operator $j$ we can measure its electric charge, which happens to be $e$. Therefore, $V_{e}$ is actually an electron operator of $U(1)$ CFT compactified on a circle with radius $R=\sqrt{m}$.  Additionally, by choosing $\alpha=\frac{1}{\sqrt{m}}$ and computing the correlation function between two of them, exchanging them counterclockwise  we obtain $\theta=\frac{\pi}{m}$ for their fractional statistics. The charge of this operator can also be measured by looking at its OPE with the current operator and turns out to be $e/m$. Accordingly, $e^{i\frac{\varphi}{\sqrt{m}}}$ represents the Laughlin quasiparticle (anyon) operator.

\subsection{Chiral algebra of the Laughlin-type states}
In this section we wish to study the operator content of the Abelian FQH/FTSC states, through the so-called chiral algebra. The conformal algebra always contains the Virasoro algebra, which is given by a central charge $c$. The edge theory of the Abelian FQH/FTSC states at $\nu=1/m$ is described by a CFT with $c=1$.  The Viraso algebra at $c=1$ allows for an infinite number of primary fields of the form $V_{\alpha}j^{n}$ with arbitrary real $\alpha$ and integer $n$ numbers, where $V_{\alpha}=e^{i\alpha \varphi}$ and $j=i\partial_{z}\varphi$. In the physical systems, however, we always have a finite number of primary fields. One way to obtain a finite number of primary fields is to extend the Virasoro algebra. A particular example is through the chiral algebra where we choose one of the primary fields as well as the current operator $j$ as the generators of the algebra and find all the primary fields that have nonsingular OPE with them. The resulting operators will form a representation of the chiral algebra. For example, in the $c=1$ we can take the electron operator $V_{e}=\exp\para{i\sqrt{m}\varphi\para{z}}$, $V_e^\dag$, and $j$ with scaling dimensions $m/2$, $m/2$ and, $1$, respectively, as the generating operators. The OPE of the current operator with vertex operators $V_{\alpha}$ is always nonsingular. Now let us consider the OPE of $V_{e}$ with vertex operators. We have 

\begin{eqnarray}
&&V_{e}\para{z}V_{\alpha}\para{w}\sim \frac{1}{\para{z-w}^{\sqrt{m}\alpha}}.
\end{eqnarray}   
In order to get a nonsingular OPE, $\sqrt{m}\alpha$ should be an integer. Therefore, every $V_{\frac{k}{\sqrt{m}}}$ operator where $k\in {Z}_{m}$ is also allowed in the chiral algebra. These operators create excitations in the $\nu=1/m$ Laughlin states. The fusion algebra of this chiral algebra is pretty simple: $V_{k/\sqrt{m}}\times V_{l/\sqrt{m}}=V_{\frac{k+l~{\rm mod}~ m}{\sqrt{m}}}$, which represents a ${Z}_{m}$ algebra. 

The chiral algebra of the free boson theory compactified on a circle with $R=\sqrt{m}$ radius can be further expanded by choosing the bosonic operators $e^{\pm i\para{2\sqrt{m}\varphi}}$ and $j$ as the generators of the chiral algebra. This alternative choice turns out to be very useful in studying the operator content of the $U(1)/Z_2$ orbifold theory description of the superconductors. The intuition is that the charge of the Cooper pair is $2$ and they are considered as the boundstate of two electrons. On the other hand, $e^{i\para{2\sqrt{m}\varphi\para{z}}}\sim \lim_{w\to z} V_{e}\para{z}V_{e}\para{w}$. Following the procedure we used above, we find the following operators as the allowed primary fields:

\begin{eqnarray}
&V_{k/2\sqrt{m}}\para{z}=e^{i\frac{k}{2\sqrt{m}}\varphi\para{z}}~,~ h_{k}=\frac{k^2}{8m}~,~k \in {Z}_{4m}.~&
\end{eqnarray}   
Form the above relation, it turns out that the previous chiral algebra is a sub-algebra of the new one. The operators that we encountered before and that represent the Laughlin state correspond to $k=2l$ and $l\leq m$. The new primary fields with $k=2l$, where $m< l \leq 2m$, were identified with $k-2m$ primary fields in the previous case ($Z_m$ algebra). The primary fields with $k=2l+1$ are all odd under the $\varphi\to \varphi+2\pi\sqrt{m}$ symmetry transformation of the original free chiral boson theory, so they can be removed form the theory at the end. The resulting theory can be easily shown, which is modular invariant, thus physical. The fusion rules for this chiral algebra reduces to a ${Z}_{4m}$ one. In general, for a chiral boson theory compactified on a circle with a rational radius of $\frac{R^2}{2}=\frac{p}{p'}$ form where $p$ and $p'$ are coprime (i.e., they their greatest common divisor is one), the maximal chiral algebra is generated by $j$, and $e^{\pm i\sqrt{2N}\varphi}$, where $N=pp'$, whose corresponding fusion algebra reduces to ${Z}_{2N}$ group.\cite{Vafa_1,DiFrancesco_1,PC_1} This chiral algebra of the $c=1$ free boson theory is referred to as $\mathcal{A}_{N}$.  As a sanity check, let us consider $R=\sqrt{m}$ theory that represents the edge state of a $\nu=1/m$ Laughlin state, $\frac{R^2}{2}=\frac{m}{2}$. Since $m$ is odd $\para{m,2}=1$, then $N=2m$. Accordingly, the maximal chiral algebra is given by $\mathcal{A}_{2m}$ and the fusion algebra reduces to ${Z}_{4m}$. 

\subsection{Vortices and the twist operators}
Imagine a superconducting state with a number of vortices. Because superconductor is a condensate of Cooper pairs with charge $q=2e$ , it allows for the vortices with a magnetic flux half of the quantum flux i.e., $\Phi_0=\frac{hc}{2e}=\pi/e$ in the unit where $\hbar$=c=1. On the other hand, the Aharanov-Bohm effect tells us that electrons moving around a vortex will pick up a minus sign. In other words, having a vortex at point $\vec{R}$ requires the introduction of a branch cut in the superconductor such that electrons change sign by passing through it. This means a vortex can be imagined as a topological defect in the superconductor. Now let us formulate this is in a more elegant way. We wish to consider an operator $\sigma \para{\vec{R}}$ referred to as the {\em twist operator} which creates a vortex at point $\vec{R}$. If we take an electron and turn it around this twist operator, the electron operator's sign changes. Within the bulk-boundary CFT theory correspondence conjecture this means we need an operator whose OPE with electrons is of the following form:
\begin{eqnarray}
&&V_{e}\para{z}\sigma\para{w}=\frac{\sigma\para{w}}{\para{z-w}^{1/2}}.
\end{eqnarray}
Going around the twist field $\sigma\para{w}$, the electron operator picks up a minus sign. Therefore, the twist operator can change the boundary condition on the electrons from periodic to antiperiodic and vice versa along space and/or time directions. Such a twist operator is quite well studied in the context of the Ising CFT and physically corresponds to the spin order and disorder fields.\cite{Ginsparg_1991, DiFrancesco_1} 

\subsection{Edge theory of the non-Abelian FTSCs: $U(1)/Z_2$ orbifold CFT}
As we discussed before, it is known that the maximal chiral algebra of a free boson compactified on a circle with $\frac{R^2}{2}=\frac{p}{p'}$ with $\para{p,p'}=1$ is given by $\mathcal{A}_{N}$, where $N=pp'$.\cite{Vafa_1,DiFrancesco_1,PC_1} This algebra is generated by the bosonic operators: $j$, and $e^{\pm i\sqrt{2N}\varphi}$. Accordingly, the primary fields allowed in this chiral algebra are $e^{i\frac{k}{\sqrt{2N}}\varphi}$, where $k\in Z_{2N}$, in addition to the current operator $j=i\partial_{z}\varphi$.  Therefore, the maximal chiral algebra of a free chiral bosons compactified on a circle with $R=\sqrt{m}$ that describes an Abelian FQH (or an Abelian FTSC) state at $\nu=1/m$ filling, is given by $\mathcal{A}_{2m}$. The free boson edge theory enjoys a discrete $Z_2$ symmetry which acts on the $U(1)$ boson field in the following way:
\begin{eqnarray}
&&\mathcal{R}:\sd \varphi \to -\varphi,
\end{eqnarray}
or, equivalently, $\mathcal{R}:~~j\to -j$. The action of the $Z_2$ group $\mathcal{R}$ divides the primary fields of the $\mathcal{A}_{2m}$ chiral algebra into the invariant and noninvariant groups: $\phi_{k}=\cos\para{k\varphi/2\sqrt{m}}$, and  $\tilde{\phi}_{k}=\sin\para{k\varphi/2\sqrt{m}}$. On the other hand, as we discussed in the previous section, for a range of parameters adding superconductivity to the system can gap out half of the degrees of the freedom corresponding to primary fields of the edge theory and keep the rest gapless, meaning either the invariant or the noninvariant part of the primary fields should be taken. As a result, the electron operator can be taken as the $\cos\para{\sqrt{m}\varphi}$, which is even under the $Z_2$ orbifold theory.  

The above argument suggests that  we should look for a chiral algebra generated by $j$, and $\cos\para{\sqrt{2N}\phi}$ ($N=2m$). The resulting CFT is known as the $U(1)/Z_2$ orbifold theory that is achieved by moding out the discrete symmetry $\mathcal{R}$ from the free boson edge CFT.  Moreover, the chiral algebra of this orbifold theory is given by $\mathcal{A}_{N}/Z_2$. 

The orbifold theory is known to be relevant in other topological states of matter as well. In Refs. \cite{PC_0,PC_1,PC_2}, Brakeshli and Wen studied a bilayer FQH state with a strong interlayer repulsion. In the absence of strong repulsion, the edge theory of the bilayer FQH state is given through two free chiral boson fields $\varphi_{1}=\frac{\varphi_{+}+\varphi_{-}}{2}$ and $\varphi_{2}=\frac{\varphi_{+}+\varphi_{-}}{2}$ and thus enjoys a $U(1)\times U(1)$ symmetry corresponding to $j_{+}=i\partial_z\varphi_{+}$ and $j_{-}=i\partial_z\varphi_{-}$ conservation. A discrete symmetry of the edge theory corresponds to the exchange of the two layers: $\mathcal{R}: \varphi_{-}\to -\varphi_{-}$. In the  $(m,m,0)$ bilayer FQH, the electron operator in each layer can be represented as $\psi_{i}=e^{i\sqrt{m}\varphi_{i}}$, where $i=1,2$. By increasing the interlayer repulsion, it is energetically favorable for the quasiparticles to reside in the same layer. This causes only the even combination of electron operators under the $\mathcal{R}$ symmetry to remain gapless and the odd combination becomes gapped. Consequently, the edge CFT is generated by $\frac{\psi_{1}+\psi_{2}}{2}=e^{i\sqrt{m}\varphi_{+}/2}\cos\para{\sqrt{m}\varphi_{-}/2}$. Therefore, the edge theory is given by the free $U(1)$ theory, corresponding to $\varphi_{+}$, and a $U(1)/Z_2$ for the the $\varphi_{-}$ part, thus a $U(1)\times U(1)/Z_2$ orbifold CFT. As in our example, the orbifold theory allows for the twist fields to show up, resulting in a non-Abelian state. In a different study, Barkeshli and Qi have shown that Chern insulators with higher Chern numbers can provide a physical playground to realize a similar physics.\cite{C_7}  More recently, Barkeshli {\em et al}. have studied the topological order and projective non-Abelian properties of these types of models.\cite{Barkeshli_2012_1}

\subsection{Operator content of the $U(1)/Z_2$ orbifold CFT}

To obtain the chiral algebra representing the non-Abelian FTSCs, we should mod out the $Z_2$ subgroup of the CFT that acts on the current operator as $\mathcal{R}:~~j\to-j $. Moding the $Z_2$ symmetry has two effects. First, the resulting primary fields should be either symmetric or antisymmetric under the orbifold action. Second, it allows new primary fields to show up in the chiral algebra. These new primary operators are twist fields. The OPE of any primary field in chiral algebra with the generators of the chiral algebra, i.e., $j$ and $\cos\para{\sqrt{2N}\varphi}$, should be nonsingular. The OPE of the twist fields with the $\cos\para{\sqrt{2N}\varphi}$ vertex operator is nonsingular. However, its OPE with the current operator $j=i\partial_{z}\varphi$ has a branch cut, which means taking the current operator around the twist field changes the sign of the current operator leading to a singular behavior of the OPE. For the unorbifold theory with the full $U(1)$ symmetry, this is not acceptable and as a result twist fields are absent in the chiral algebra. However, by moding the $Z_2$ symmetry generated by $\mathcal{R}$, the twist fields are allowed because $-j\para{z}=\mathcal{R}^{-1}j\para{z}\mathcal{R}\sim j\para{z}$. In other words, the current operator is invariant after going around the twist fields up to the action of the $Z_2$ group. Therefore, the chiral algebra of the $\mathcal{A}_{N}/Z_2$ theory has to include twist fields as well. It is these twist fields that make the orbifold theories significantly richer and more exciting than the original $\mathcal{A}_{N}$ chiral algebra associated with the Abelian FTSC.

There are two kinds of twist fields in the $U(1)/Z_2$ orbifold theories: $\sigma_1$ and $\sigma_2$ fields with conformal dimension $h_{\sigma}=1/16$ and $\tau_1$ and $\tau_2$ fields with conformal dimension $h_{\tau}=1/16+1/2=9/16$. The latter twist fields are related to the first one through the following relation:

\begin{eqnarray}
&&j\para{z}\sigma_{i}\para{w}\sim \frac{\tau_{i}\para{w}}{\para{z-w}^{1/2}},
\end{eqnarray}
which is consistent with the fact that $\sigma_{i}\para{w}$ twist operators change the sign of the $j\para{z}=i\partial_{z}\varphi\para{z}$ field through $2\pi$ rotation around $w$. As was stated before, twist operators are allowed by the chiral algebra $\mathcal{A}_{N}/Z_2$ since under the action of the twist field on the current operator, it is invariant up to the action of the $Z_2$ group (more precisely $j\equiv -j$ in the $Z_2$ orbifold theory). 

The generators of the chiral algebra are the current ($j$) and $\cos\para{\sqrt{2N}\varphi}$ operators  ($N=2m$). We now consider the rest of primary fields that have nonsingular OPE with these generators other than twist fields. It can be shown that the following set of operators has nonsingular OPE with these operators.

\begin{eqnarray}
&&\phi_{k}=\cos\para{\frac{k}{\sqrt{2N}}\varphi}=\cos\para{\frac{k}{2\sqrt{m}}\varphi}.
\end{eqnarray}
which are simply the $Z_2$-invariant part of the primary fields of the $\mathcal{A}_{2m}$ chiral algebra that we encountered before.  There are two fermion operators in the $\mathcal{A}_{2m}/Z_2$ chiral algebra among which we take the $Z_2$-invariant one as the electron operator. They are 
\begin{eqnarray}
&&V_{e}=\phi_{N}^{1}=\cos\para{\sqrt{\frac{N}{2}}\varphi}=\cos\para{\sqrt{m}\varphi},\cr
&&\phi_{N}^{2}=\sin\para{\sqrt{\frac{N}{2}}\varphi}=\sin\para{\sqrt{m}\varphi}.
\end{eqnarray}

Several properties of these primary fields are summarized in table [\ref{tab1}]. The most important information one needs in any CFT is the fusion algebra of the primary fields. For the $Z_2$ orbifold theories we are concerned with here, Dijgkgraaf {\em et al}. \cite{Vafa_1} have derived their fusion rules. According to their study, for even $N$ (in our problem it is the case) the fusion algebra is as follows:

\begin{table}
\label{tab1}
  \centering
  \caption{Primary fields of the orbifold theory given by an§ $\mathcal{A}_{N}/Z_2$ chiral algebra with $N=2m$, and the corresponding conformal (scaling) and quantum dimensions.}\
  \begin{tabular}{c c c  c}
  \hline
  \hline
  $\sd \sd$& \sd ${\rm CFT~primary}$\sd&\sd${\rm Conformal}$\sd~~&~~\sd${\rm Quamtum}$\\
  $\sd\sd$& \sd${\rm field}$\sd&\sd${\rm dimension}$\sd~&~\sd${\rm dimension}$\\
  \hline
   ${\rm \bf 0}$& $1$&$0$&$1$\\
   ${\rm \bf 1}$&  $j$&$1$&$1$\\
   ${\rm \bf 2}$&  $\phi_{2m}^{i}$&$m/2$&$1$\\
${\rm \bf 3}$& $\phi_{k}$&$k^2/\para{8m}$&$2$\\
 ${\rm \bf 4}$&    $\sigma_{i}$&$1/16$&$\sqrt{2m}$\\
  ${\rm \bf 5}$&   $\tau_{i}$&$9/16$&$\sqrt{2m}$\\
\hline
\hline
  \end{tabular}
\end{table}

\begin{eqnarray}\label{fus_1}
&& \phi_{k}\times \phi_{k'}=\phi_{k+k'}+\phi_{k-k'}\sd \para{k'\neq k, N-k}, \cr
&& \phi_{k}\times \phi_{k}=1+j+\phi_{2k},\cr
&& \phi_{N-k}\times \phi_{k}=\phi_{2k}+\phi_{N}^{1}+\phi_{N}^{2}.
\end{eqnarray}
The above fusion rules can be verified by interpreting $\phi_{k}$ fields as $\cos\para{k\varphi/\para{\sqrt{2N}}}$.  Also, the fusion rules involving $j$, $\phi_{N}^{1}$ and $\phi_{N}^{2}$ operators are:
\begin{eqnarray}\label{fus_2}
&&j\times \phi_{k}=\phi_{k},\cr
&& j\times j=1, \cr
&& \phi_{N}^{i}\times \phi_{N}^{i}=1,\cr
&& \phi^{1}_{N}\times \phi_{N}^{2}=j,
\end{eqnarray}
and, most importantly, the fusion rules involving the twist operators is given by the following formulas:
\begin{eqnarray}\label{fus_3}
&& \sigma_{i}\times \sigma_{i}=1+\phi_{N}^{i}+\sum_{k~{\rm even}}\phi_{k}, \cr
&& \sigma_{1}\times \sigma_{2}=\sum_{k~{\rm odd}}\phi_{k},\cr
&& j\times \sigma_{i}=\tau_{i},\cr
&&\phi_{k}\times \sigma_{i}=\sigma_{i},\cr
&&\phi_{N}^{i}\times \sigma_{i}=\sigma_{i}.
\end{eqnarray}

The fusion rules for the twist fields $\tau_{i}$ can also be read off the $\tau_{i}=j\times \sigma_{i}$ formula. The above relations suggest that the $\mathcal{A}_{2m}/Z_2$ chiral algebra supports $2m+7$ primary fields, among which $\phi_{N}^{1}$, $\phi_{N}^{2}$, and $j$ are Abelian fields and $\phi_{k}$'s are non-Abelian primary fields with a quantum dimensions equal to $2$ and the non-Abelian twist fields with a quantum dimensions equal to $\sqrt{N}$. Now let us make comments on the above fusion algebra. First of all, as we stated before, $\phi_{2m}^{1}$ field with conformal dimension $h=m/2$ is a fermion field and invariant under under the $\mathcal{R}$ symmetry. We choose it to represent the electron operator. The above fusion rule between an electron operator and a twist field $\sigma_{1}$ along with their conformal dimensions suggest the following OPE (and a similar relation for the $\tau_{1}$):

\begin{eqnarray}
V_{e}\para{z}\sigma\para{w}=\frac{\sigma\para{w}}{\para{z-w}^{m/2}}.
\end{eqnarray}  

This relation is consistent with the fact that taking an electron around the twist field flips the sign of the electron operator and changes the boundary conditions. Second, $\phi_{2l}=\cos\para{\frac{l}{\sqrt{m}}}$ operators represent the $\mathcal{R}$ invariant non-Abelian anyon operators with conformal dimensions $h_{2l}=l^2/2m$. The OPE of these anyon fields with the twist fields $\sigma_{1}$ is (and a similar relation for the $\tau_{1}$)

\begin{eqnarray}
\phi_{2l}\para{z}\sigma\para{w}=\frac{\sigma\para{w}}{\para{z-w}^{l^2/2m}},
\end{eqnarray}
which means taking an anyon operator around the twist field will change its phase by $\theta_{2l}=l^2\pi/m$. We use this important observation in studying the zero modes of the anyon operators.


\section{Topological order of the non-Abelian FTSCs}
In this section, we study different properties of the non-Abelian excitations. We prove the existence of the zero modes bound to vortices, compute their quantum dimensions and derive the non-Abelian part of the braid matrix that acts on the degenerate ground state by braiding fractionalized Majorana zero modes.

\subsection{Mode expansion and zero modes}

In this section,  we study the physical effects of the twist fields and show that they allow for the existence of the fractionalized Majorana zero modes. First of all, let us mention the following mode expansion for the primary fields. For an arbitrary chiral primary field $\Phi\para{z}$ with conformal dimension $h_{\Phi}$ we can write the following mode expansion:

\begin{eqnarray}
\Phi\para{z}=\sum_{n=n'-\frac{p}{q}} \frac{\alpha_{n}}{z^{n+h_{\Phi}}}\sd n',p,q\in Z.
\end{eqnarray}

In order to determine the allowed values for $p$ and $q$ integers, we need to know the behavior of the $\Phi\para{z}$ field under the twist, i.e., the value of the $\theta$ in $\Phi\para{e^{2\pi i}z}=e^{-i\theta}\Phi\para{z}$. On the other hand, the above mode expansion suggests that $\Phi\para{e^{2\pi i}z}=e^{-i\para{h_{\Phi}-p/q}2\pi}\Phi\para{z}$. Therefore, $\frac{p}{q}=h_{\Phi}-\frac{\theta}{2\pi} ~{\rm mod}~ 1$. If $p/q=0$, then it means $\alpha_{0}$ is allowed in the system and we say that the $\Phi$ field supports a zero mode. For physical systems we {\em usually} have periodic boundary conditions on fields, i.e., $\theta=0$, and therefore for the primary fields with rational conformal dimension zero mode is absent. For example, in the Laughlin FQH state at $\nu=1/m$ filling fraction, $\frac{p}{q}=\frac{m}{2}-\frac{0}{2\pi} ~{\rm mod}~ 1=1/2$. This means $n=m+1/2$, and accordingly, there is no electron zero mode.  An interesting question that arises is what can cause a twisted boundary condition (i.e., $\theta \neq 0$) on the $\Phi\para{z}$? The answer is simply the twist fields. A physical realization of the twist field is the flux insertion (vortex) that through the Aharanov-Bohm effect can change the boundary condition. For example, a single half vortex with $\phi_{0}/2$ magnetic flux can change the boundary condition on electrons from a periodic to an antiperiodic one.  For such a boundary condition, $\frac{p}{q}=\frac{m}{2}-\frac{\pi}{2\pi} ~{\rm mod}~ 1=0$. Consequently, the zero mode is now allowed. This is, for example, the reason why zero modes are bound to the vortices in $p+ip$ superconductors.

As the fusion rules of the $U(1)/Z_2$ orbifold theory suggest (see Eq. [\ref{fus_3}]), fusing two twist fields (vortices) of the same kind can generate $\phi_{2k}$ primary fields. Now let us take the anyon primary field with the least conformal dimension i.e. $\phi_{2}=\cos\para{\frac{\varphi\para{z}}{\sqrt{m}}}$ among them. The mode expansion of this primary field whose conformal dimension is $\frac{1}{2m}$ reads as follows:

\begin{eqnarray}
\phi_{2}\para{z}=\sum_{n=n'-\frac{p}{q}} \frac{\alpha_{n}}{z^{n+1/{2m}}}\sd n',p,q\in Z.
\end{eqnarray}
In the usual periodic boundary condition, $p/q=1/m$, and therefore $n=n'-1/(2m)$, so the zero-mode operator, $\alpha_{0}$, is absent. Now imagine we have put a twist field corresponding to the half vortices at $w=0$. Taking the $\phi_{2}\para{z}$ field around the twist field will result in $\theta=\frac{\pi}{m}$. As a result, in this case $\frac{p}{q}=\frac{1}{2m}-\frac{\theta}{2\pi} ~{\rm mod}~ 0$. Accordingly, $n \in Z$ and the zero-mode operator $\alpha_0$ exists in the mode expansion of the $\phi_2\para{z}$ field in the presence of a twist field at $w=0$ (or by flux insertion at $w=0$).

Let us consider the fusion algebra between $\phi_{k}$ and $\phi_{2N-k}$ fields, which is $\phi_{k}\times \phi_{2N-k}=1$. This means $\phi_{2N-k}$ is the conjugate of the $\phi_{k}$; i.e., their fusion gives the vacuum. Thus, we represent the $\phi_{2N-k}$ as $\phi_{k}^\dag$. In particular, take the anyon operator, $\phi_{2}$, whose conjugate field is $\phi_{2\para{m-1}}$ with a conformal dimension equal to $\frac{\para{2m-1}^2}{2m}$. According to their fusion, the OPE between these fields is given by the following formula  (up to a constant number):

\begin{eqnarray}
&&\phi_{2}^\dag\para{z}\phi_{2}\para{w} \sim \frac{1}{\para{z-w}^{2\para{m-1}+1/m}},\cr
&&\phi_{2}\para{w}\phi_{2}^\dag\para{z}\sim \frac{1}{\para{w-z}^{2\para{m-1}+1/m}},
\end{eqnarray}
from which we obtain the following commutation relation:
\begin{eqnarray}
&&\phi_{2}^\dag\para{z}\phi_{2}\para{w}=e^{-i\pi/m}\phi_{2}\para{w}\phi_{2}^\dag\para{z}.
\end{eqnarray}

Using the mode expansion, the commutation relation between the modes can be achieved, which is
\begin{eqnarray}
&&\alpha_{k}^\dag \alpha_{l}=e^{-i\pi/m}\alpha_{l}\alpha_{k}^\dag.
\end{eqnarray}
Another important property of the $\phi_{2}$ field is that fusing it $2m$ times with itself gives the identity, i.e., $\phi_{2}^{2m}\sim 1$. This property has an important consequence for the zero-modes. To see that we first mention that modes can be written in terms of the following contour integrals:

\begin{eqnarray}
&&\alpha_{k}=\oint_{c} \frac{dz}{2\pi} \phi_{2}\para{z} z^{k+1/\para{2m}-1},
\end{eqnarray}
where $c$ encloses the origin.  Using this relation, it can be shown that

\begin{eqnarray}
&&\alpha_{0}^{2m}=\alpha_{0}^\dag{}^{2m}=1.
\end{eqnarray}
The above relation suggests that $\alpha_0$ is the generalization of the Majorana zero $\psi_0$ mode for which $\psi_{0}^{2}=1$. In this problem $\alpha_{0}^{m}=\alpha_{0}^\dag{}^{m}$ acts like a Majorana zero mode and for $m_0>1$ we refer to the zero mode operator $\alpha_{0}$ as the {\em fractionalized Majorana zero mode} for obvious reasons.

\subsection{Quantum dimension of the non-Abelian excitations}

One of the features of the topological order is the GSD. This degeneracy is related to the so-called quantum dimension of the zero modes. A zero mode, as its name implies, excites quasiparticles with zero energy. In other words, by applying a zero-mode operator on the ground state we yield a new state with the same energy. The degeneracy of the ground state is a topological index of the system which is robust against perturbations. Clearly, the ground state(s) of a Hamiltonian depends on the boundary conditions. The existence of some fields can change the boundary condition and as a result the GSD.  For example, as we mentioned in detail before, a twist operator at the $\vec{R}_1$ point exchanges the boundary condition on electrons moving around it and gives the possibility of the zero modes.  One way to define the quantum dimension of an operator, in this case twist fields, is to consider $n$ of such operators at isolated points $\para{\vec{R}_1, ..., \vec{R}_{n}}$. Then compute the GSD with respect to this constraint. The GSD usually grows with the number of such fields. Accordingly, the quantum dimension of these operators is defined as follows:

\begin{eqnarray}
\ln d=\frac{\ln GSD}{n}\sd n\to \infty.
\end{eqnarray}

There are several other ways to define the quantum dimension of an operator, but we stick to the above definition for our purposes in this paper.  Now let us consider an FTSC on a torus (or an infinite plane) with two vortices at $\vec{R}_1$ and $\vec{R}_2$ in it. The vortices can be modeled by putting $\sigma_1$ twist fields at these points. As we argued before, twist fields allow the existence of zero mode operators bound to (localized at) twist operators. This bound state can be justified in an intuitive way. Naively, a vortex means a circular puncture with radius $L$ (of the order of the correlation length of Cooper pairs) centered at $\vec{R}_1$. We  expect an edge state propagating along the boundary of the puncture and localized in the perpendicular direction. The reason for this expectation is that inside the puncture is a trivial state while outside it is an FTSC. The chiral anyons (corresponding to the $\phi_2$ field ) that propagate along the boundary have energies of the form $E=v_{\rm F}k$, where $k$ is their momentum and $v_{\rm F}$ represents the Fermi velocity. However, since $L$ is finite, the momentum of the quasiparticles is quantized as $k_{n}=\frac{2\pi n}{L}$, where $n$ is an integer due to the fact that the twist field (vortex) allows for the zero mode.  Therefore, level spacing is $\Delta E\propto 1/L\to \infty$, meaning that only the zero mode ($n=0$) can be observed at that point. Therefore, twist fields at $\vec{R}_{1}$ allows for the fractionalized Majorana zero mode, $\alpha_0$, at that point and similarly does the twist field at $\vec{R}_2$. Let $\gamma_{1}$ and $\gamma_{2}$ operators denote zero modes at $\vec{R}_1$ and $\vec{R}_2$. They follow the following algebra:

\begin{eqnarray}
&&\gamma_{1}^\dag \gamma_{2}=\exp\para{-i\pi/m}\gamma_{2}\gamma_{1}^\dag \sd,\sd \gamma_{1}^{2m}=\gamma_{2}^{2m}=1.
\end{eqnarray}
Using the above algebra we can show that $\gamma_{1}^\dag \gamma_{2}$ operator satisfies the relation

\begin{eqnarray}
&&\para{\gamma_{1}^\dag \gamma_{2}}^{2m}=\exp\para{-i\pi\para{2m+1}}=-1,
\end{eqnarray}
which defines a $2m$ state with the following eigenvalues:
\begin{eqnarray}
&&e^{i\pi/(2m)}\gamma_{1}^\dag \gamma_{2}\ket{q}=\exp\para{i\pi q/m}\ket{q}.
\end{eqnarray}
Therefore, we can consider $e^{i\pi/(2m)}\gamma_{1}^\dag \gamma_{2}$ as the generalization of the familiar fermion parity operator for Majorana fermions. Hence, we call it the $Z_{2m}$ character of the ground state. To understand the nature of the degenerate ground states, the following relations that can be easily proven will be found useful:
\begin{eqnarray}
&&\gamma^\dag_{1}\ket{q}=\exp\para{-i\pi\xi_1 \para{q-1/2}/m}\ket{q+1 ~{\rm mod}~ 2m},\cr
&& \gamma_{1}\ket{q}=\exp\para{i\pi\xi_1 \para{q-1/2}/m}\ket{q-1 ~{\rm mod}~ 2m},\cr
&&\gamma^\dag_{2}\ket{q}=\exp\para{-i\pi\xi_2 \para{q-1/2}/m}\ket{q+1 ~{\rm mod}~ 2m},\cr
&&\gamma_{2}\ket{q}=\exp\para{i\pi \xi_2\para{q-1/2}/m}\ket{q-1 ~{\rm mod} ~2m},
\end{eqnarray}
where $\xi_2-\xi_1=1$, and $\xi_2$ can be any real number. The ambiguity in the value of the free parameter $\xi_2$ indicates that only quadratic combinations of the form $\gamma_{1}^\dag \gamma_{2}$ are unambiguous, and thus physical and measurable. However, the value of the $\xi_2$ does not affect the quantum dimension of the fractionalized Majorana zero modes ($\gamma_{i}$ operators) as well as their projective braid statistics. Therefore, we fix the gauge and assume $\xi_2=1$, and $\xi_1=0$ as of now. With this gauge convention, the ground state $\ket{q}$ can be constructed in either of the following ways:
\begin{eqnarray}\label{gs_1}
&&\ket{q}=\gamma^\dag_{1}{}^{q}\ket{0}=\exp\para{i\pi/2m\para{q^2-2q}}\gamma_{2}^\dag{}^{q}\ket{0}.
\end{eqnarray}
The above relations play a key role in our paper. To see how powerful the CFT methods is, note the fusion rule given for two twist fields in Eq. [\ref{fus_3}].  It simply means that taking every two twist fields (vortices), their composite state results in different fusion channels which is a signature of their non-Abeilan nature. The final state can be either vacuum or a state with one of the anyon operators $\phi_{2}$, $\phi_{4}$, ..., and $\phi_{4m-2}$. Note that the zero mode of the $\phi_{2k}$ operator can be created by taking the $k$-th power of the fractionalized Majorana zero mode of the $\phi_{2}$ i.e. $\gamma_{i}^{k}$ ($i$ can be either $1$ or $2$ since we have fused twist fields at $R_1$ and $R_2$), which can be easily proven using the fusion rules for $\phi_{2l}$ operators. Therefore, fusing every two twist operators results in one of the following states $\left\{\ket{0}, \gamma_{i}^\dag\ket{0}, \gamma_{i}^{\dag}{}^{2}\ket{0}, ..., \gamma_{i}^\dag{}^{2m-1}\ket{0}\right\}$. As a concluding remark, remember that fractionalized Majorana zero-mode operator $\gamma_{i}$ is bound to $\sigma_{1}\para{R_{i}}$. Therefore, fusing two twist fields can be viewed as considering a $\gamma_{1}^\dag \gamma_{2}$ operator, which happens to define a $2m$ degenerate states. So the results of the CFT are consistent with the direct calculation.  

The main result of our discussion so far is that by taking any two vortices, we can define the $Z_{2m}$ character $e^{i\pi/(2m)}\gamma_{i}^\dag \gamma_{j}$ that takes values in $Z_{2m}$ and as a result defines $2m$ degenerate states. Accordingly, taking $n$ vortices, we can define $n/2$ of such $Z_{2m}$ characters (remember that in a compact space there are always an even number of vortices) and results in the $\para{2m}^{n/2}$ degenerate states. However, we should note that the number of electrons in a superconductor is conserved mod 2. Each electron operator is formed of $m$ anyon operators ($\phi_2$ fields) in $\mathcal{A}_{2m}/Z_2$ chiral algebra. So the number of anyons is conserved mod $2m$. This generalized parity of anyons can be computed through the $Z_{2m}$ characters. It is indeed $\chi=\prod_{i=1}^{n/2}e^{i\pi/(2m)}\gamma_{2i-1}^\dag\gamma_{2i}$. The conservation of the $\chi$ operator reduces the number of degenerate states by a factor of $1/(2m)$, since one of the $Z_{2m}$ characters is fixed by the value of the $\chi$ and the rest of $Z_{2m}$ characters. Altogether, we reach the following conclusion:

\begin{eqnarray}
GSD\para{n}=\para{2m}^{n/2-1} \Rightarrow d_{\sigma}=\sqrt{2m}.
\end{eqnarray} 
which agrees with the result of the CFT.

\subsection{Projective braiding statistics of  the fractionalized Majorana zero modes}
Braiding means an adiabatic evolution of the ground state in space-time, after which two quasiparticles exchange their position. For example, exchanging the position of two electrons causes the change of the ground state wave function by $\pi$. In the FQH state at $\nu=1/m$, exchanging the position of two anyons counterclockwise changes the phase of the ground state wave function by $\pi/m$ phase factor. For non-Abelian quasiparticles, however, the ground state is degenerate and exchanging two quasiparticles may cause a unitary transformation of the ground state; i.e., it may mix different degenerate ground states after exchanging two non-Abelian quasiparticles (for example two Majorana zero modes) adiabatically. In the CFT, braid matrix (that acts on the ground states) is defined through the four-point correlation functions. Take four primary fields $\Phi\para{z_1}$, $\Phi\para{z_2}$, $\Phi\para{z_3}$, and $\Phi\para{z_4}$. Then compute the following correlation function:

\begin{eqnarray}
\braket{\Phi\para{z_1}\Phi\para{z_2}\Phi\para{z_3}\Phi\para{z_4}}.
\end{eqnarray}
Suppose the fusion algebra of the $\Phi$ primary fields is given by $\Phi\times \Phi=\sum_{l=1}^{d}\Omega_{l}$ for some know $\Omega_{l}$ fields. Using this fusion algebra, there are several  ways to compute the four-point correlation function. One way is to fuse $\Phi_{1}$ with $\Phi_{2}$ and $\Phi_{3}$ with $\Phi_{4}$ in the $l$ channel and sum over different channels. For each particular channel let us call the result $\mathcal{F}_{12}^{34}\para{l,z}$, where $z=\frac{z_{12}z_{34}}{z_{13}z_{24}}$ and $z_{ij}=z_i-z_j$. The second way is  to fuse $\Phi_{1}$ with $\Phi_{3}$ and $\Phi_{2}$ with $\Phi_{4}$ in the $l'$ channel and sum over different $l'$ channels. Again for each particular channel let us call the result $\mathcal{F}_{13}^{24}\para{l',1/z}$.  The consistency check requires $\mathcal{F}_{12}^{34}\para{l,z}$ to be a linear combination of $\mathcal{F}_{13}^{24}\para{l',1/z}$ functions say:
\begin{eqnarray}
\mathcal{F}_{12}^{34}\para{l,z}=\sum_{l'}\para{B_{23}}_{l}^{l'}\mathcal{F}_{13}^{24}\para{l',1/z}.
\end{eqnarray}
$B_{23}$ is called the braid matrix associated with braiding $\Phi_{2}$ and $\Phi_{3}$, which acts like a unitary transformation on the degenerate ground states formed of four $\Phi$ field insertions.  Let us go back to our problem. Upon vortex (the dual of the edge CFT's twist fields in the bulk of the FTSC) insertion, a fractionalized Majorana zero mode is created at that point. Instead of braiding zero modes, we can instead braid vortices that obviously braid the zero modes bound to them as well. It should be emphasized that the braid matrix can be decomposed in the Abelian (phase factor) and non-Abelian parts. The Abelian part depends on the microscopic details of the system and the interaction between superconducting vortices. Since it does not affect the non-Abelian part of the braid matrix,\cite{Rev_1} we neglect the Abelian part of the braid statistics in this paper and refer to the remaining non-Abelian part as the projective braid matrix.\cite{Barkeshli_2012_1} The calculation of the projective braid matrix from the conformal blocks using the above recipe is straightforward but tedious. In this section, we employ a somewhat easier but related method that uses the $S$ matrix to compute the non-Abelian part of the $B_{23}$ matrix. To this end, we first compute $B_{12}$, which acts on the ground state upon exchanging twist fields $\sigma_1$ and $\sigma_2$ (as well as the zero modes bound to them).  To compute $B_{12}$ we can actually forget about other twist fields and consider vortices $1$ and $2$ only. From the fusion rule given for two twist fields in Eq. [\ref{fus_3}],  we can write the OPE between $\sigma_1$ and $\sigma_2$ fields (up to constant coefficients). We have:

\begin{eqnarray}\label{OPE_10}
&& \sigma_{1}\para{z}\sigma_{1}\para{w}=\cr
&&\frac{1}{\para{z-w}^{1/8}}\left[\sum_{q=0}^{2m-1}\para{z-w}^{q^2/\para{2m}}\phi_{2q}\para{\frac{w+z}{2}}\right].~~~
\end{eqnarray}
Remember that $\phi_{2m}=\phi_{2m}^{1}=V_{e}$ and $\phi_{0} \equiv \phi_{4m}\equiv 1$. The above formula means that taking two twist fields and fusing them (and as a result fractionalized Majorana zero modes bound to them) is ambiguous and depends on the fusion channel. The $q$th fusion channel contains $k$ fractionalized Majorana zero modes. Therefore, exchanging two twist fields does not simply give us a phase change like Abelian quasiparticles. Instead, the phase factor that we obtain depends on the fusion channel. By exchanging $z$ and $w$ [accordingly $\sigma_1\para{z}$ and $\sigma_1\para{w}$] in the $q$th channel, one can verify from the above OPE that it results in an $e^{i\pi/8}e^{-i\pi q^2/2m}$ phase factor. Forgetting the Abelian common factor in every fusion channel, i.e., $e^{i\pi/8}$, and keeping the $q$-dependent part only, we conclude that in the $q$-th channel, i.e., $\ket{q}=\gamma_{1}^\dag {}^{q}\ket{0}$ exchanging $\gamma_{1}$ and $\gamma_{2}$ as a result of exchanging twist fields, results in $e^{-i\pi q^2/2m}\ket{q}$. Therefore,

\begin{eqnarray}
&&B_{12}\ket{q}=\para{B_{12} \gamma_{1}^\dag B_{12}^\dag}^{q}B_{12}\ket{0}=\cr
&&\exp\para{-i\pi q^2/2m}\ket{q}=\exp\para{-i\pi q/m}\gamma_{2}^\dag{}^{q}\ket{0},
\end{eqnarray}
stating that the $B_{12}$ operator is diagonal in this basis. From Eq. [\ref{gs_1}], we can rewrite the above formula as:
\begin{eqnarray}\label{Braid_2}
&&B_{12}\ket{q}=\exp\para{-i\pi q/m}\gamma_{2}^\dag{}^{q}\ket{0}.
\end{eqnarray}

Since $B_{12}\ket{0}=\ket{0}$, by comparing Eqs. [\ref{gs_1}] and [\ref{Braid_2}] we find that $B_{12} \gamma_{1} B_{12}^\dag=e^{i\pi/m}\gamma_{2}$. To find the way the braid operator acts on $\gamma_{2}$ we can use a similar procedure. We can also use the fact that the $Z_{2m}$ character, i.e., the $e^{i\pi/(2m)}\gamma_{1}^\dag \gamma_{2}$ operator, is a physical and measurable quantity and cannot change after braiding $\gamma_1$ and $\gamma_2$ and keeping the rest of the system untouched. To guarantee this physical expectation, the braid operator has to act on $\gamma_{1}$ and $\gamma_{2}$ as follows:

\begin{eqnarray}
&B_{12} \gamma_{1} B_{12}^\dag=e^{i\pi/m}\gamma_{2},&\cr
&B_{12}\gamma_{2} B_{12}^\dag=e^{i\pi/m}\gamma_{2}\gamma_{1}^\dag \gamma_{2}.~~&
\end{eqnarray}

Now we wish to compute the $B_{23}$ braid matrix from the diagonal $B_{12}$ braid matrix. For this purpose, we have to consider four vortices at least. Imagine we are given four vortices at $R_{i}$ points, where $i= \{1,..,4\}$. Taking the fist two vortices, we can define a $\ket{q}_{1,2}$ state that diagonalizes the $e^{i\pi/(2m)}\gamma_{1}^\dag \gamma_{2}$ operator as well as $B_{12}$. We can do a similar job for the third and fourth vortices and define a $\ket{q'}_{34}$ state on which the $e^{i\pi/(2m)}\gamma_{3}^\dag \gamma_{4}$ and $B_{12}$ operators act diagonally. Therefore, the ground state of the whole system is $\ket{q,q'}_{\para{1,2}\para{3,4}}$. However, we should emphasize that $q$ and $q'$ are not independent as the total $Z_{2m}$ character of the system $\chi_{\rm T}=e^{i\pi/(2m)}\gamma_{1}^\dag \gamma_{2}e^{i\pi/(2m)}\gamma_{3}^\dag \gamma_{4}=e^{i\pi\frac{q+q'}{m}}$ is fixed. For simplicity, let us assume $\chi_{\rm T}=1$, so $q+q'=0~{\rm mod}~2m$. So we have:

\begin{eqnarray}
&&B_{12}\ket{q,\bar{q}}_{\para{1,2}\para{3,4}}=\exp\para{-i\pi q^2/2m}\ket{q,\bar{q}}_{\para{1,2}\para{3,4}},\cr
&& B_{34}\ket{q,\bar{q}}_{\para{1,2}\para{3,4}}=\exp\para{-i\pi q^2/2m}\ket{q,\bar{q}}_{\para{1,2}\para{3,4}}.~~
\end{eqnarray}
To find $B_{23}$ we need a basis in which $e^{i\pi/(2m)}\gamma_{3}^\dag \gamma_{2}$ and as result $B_{23}$ is diagonal. Since $\chi_{\rm T}=1$ is conserved, $e^{i\pi/(2m)}\gamma_{1}^\dag \gamma_{4}$ will be diagonal on that basis as well. In other words, we need states of the form $\ket{p,\bar{p}}_{\para{2,3}\para{1,4}}$ in which $B_{23}$ acts as $e^{-i\pi p^2/2m}$. So the procedure is simple. Through basis rotation, we can diagonalize $B_{23}$ and then we rotate back to the former bases. Let us assume there is an $S$ matrix that relates the two bases. This is actually the same $S$ matrix that we encounter in the CFT on a 2D torus that exchanges the two nontrivial cycles of the torus. Let us find the $S$ matrix by simple calculations. We have

\begin{eqnarray}
&&\ket{p,\bar{p}}_{\para{3,2}\para{1,4}}=\sum_{p}S_{p}^{q}\ket{p,\bar{p}}_{\para{1,2}\para{3,4}}.
\end{eqnarray}
The above relation suggests that $e^{i\pi/(2m)}\gamma_{3}^\dag \gamma_{2}$ acts diagonally on the $\ket{p,\bar{p}}_{\para{3,2}\para{1,4}}$ state with an eigenvalue equal to $e^{i\pi p/m}$. On the other hand, 
\begin{eqnarray}
&&e^{i\pi/(2m)}\gamma_{3}^\dag\gamma_{2}\ket{q,\bar{q}}_{\para{3,2}\para{1,4}}=\cr
&&\sum_{p}S_{p}^{q}\ket{p-1,\bar{p}+1}_{\para{1,2}\para{3,4}}=\sum_{p}S_{p+1}^{q}\ket{p,\bar{p}}_{\para{1,2}\para{3,4}}.\cr
&&
\end{eqnarray}
To have a consistent result, we should have: $S_{p+1}^{q}=\exp\para{iq\pi/m}S_{p}^{q}$ which can be solved as follows:

\begin{eqnarray}
&&S_{p}^{q}=\frac{\exp\para{ipq\pi/m}}{\sqrt{2m}}
\end{eqnarray}
To go from the $\ket{p,\bar{p}}_{\para{3,2}\para{1,4}}$ to the $\ket{q,\bar{q}}_{\para{1,2}\para{3,4}}$ basis, we can use the $S^{-1}$ matrix. It can be easily verified that $S^{-1}=S^{\dag}$. Now let use start from the $\ket{q,\bar{q}}_{\para{1,2}\para{3,4}}$ state and braid vortices $2$ and $3$. To this end, we fist use $S^\dag$ to go to the $\ket{p,\bar{p}}_{\para{3,2}\para{1,4}}$ basis. Then we act on this state by $B_{23}$, which is diagonal now. We finally rotate back to the  $\ket{q,\bar{q}}_{\para{1,2}\para{3,4}}$  basis. These steps can be seen more explicitly in the following relations:
\begin{eqnarray}
&&B_{23}\ket{q,\bar{q}}_{\para{1,2}\para{3,4}}=\sum_{q}S^\dag_{q}{}^{p}B_{23}\ket{p,\bar{p}}_{\para{3,2}\para{1,4}}=\cr
&&\sum_{q}S^{\dag}_{q}{}^{p}\exp\para{-ip^2\pi/2m}\ket{p,\bar{p}}_{\para{3,2}\para{1,4}} =\cr
&&\sum_{q,k}S^\dag_{q}{}^{p}\exp\para{-ip^2\pi/2m}S_{k}^{p}\ket{k,\bar{k}}_{\para{1,2}\para{3,4}}=\cr
&&\sum_{k}\para{B_{23}}_{k}^{p}\ket{k,\bar{k}}_{\para{1,2}\para{3,4}},\nonumber
\end{eqnarray}
which can be summarized as:
\begin{eqnarray}
&&B_{23}=S^\dag B_{12} S~,~ \para{B_{12}}_{p}^{q}=\delta_{p,q}e^{-iq^2\pi/(2m)}.~~~
\end{eqnarray}
Since $B_{23}$ is not diagonal in the $\ket{q,\bar{q}}_{\para{1,2}\para{3,4}}$ basis, it means that exchanging vortices $2$ and $3$ (or $1$ and $4$) causes a unitary non-Abelian transformation on the ground state. In other words, it will creat other degenerate ground states upon braiding vortices $2$ and $3$. In contrast, after braiding vortices $1$ and $2$ (or $3$ and $4$), the ground state  just acquires an Abelian phase which depends on the ground state though. An interesting question that arises concerns how braiding can change the ground state despite its adiabatic nature. The answer is that all the degenerate ground states share exactly the same physically measurable quantities. For example, they all have the same $Z_{2m}$ characters and the same energies. Additionally, having vortices in the system causes branch cuts. Upon braiding, some of the zero modes that bound to vortices have to cross these branch cuts. As a result, the zero modes at two vortices involved in this crossing will rearrange. This rearrangement modifies the wave function of the many-body system and the resulting wave function should be still be in the ground space, meaning that it is a linear combination of different degenerate sectors of the ground space. 

\section{Wave function of FTSCs}
As we discussed in Sec. III, according to the bulk-boundary correspondence conjecture for the chiral states, e.g., FQH states, the wave function of the bulk Hamiltonian can be written in terms of the conformal blocks of the edge-state CFT in an appropriate way. The first step is to identify the CFT that describes the edge of a chiral state. For an FTSC whose parent (unpaired) state is a Laughlin state at $\nu=1/m$, as argued in Sec. III, the edge CFT is given by the $\mathcal{A}_{2m}/Z_2$ chiral algebra of a $U(1)/Z_2$ orbifold theory. The primary fields of this CFT contain two fermion fields $\phi_{2m}^{1}=\cos\para{\sqrt{m}\phi\para{z}}$ and  $\phi_{2m}^{1}=\sin\para{\sqrt{m}\phi\para{z}}$. These are the counterpart of the Majroana fermion fields for the $m=1$ case. In the FTSC, one of these two operators become gapped and the other remains gapless. Let us assume that $\phi_{2m}^{1}$ is gapless. The wave function in the absence of vortices for a fixed number of electrons with positions at $\left\{ z_i \right\}$, where $i=\{1,..,N\}$, can be written in the following way:
\begin{eqnarray}
&&\ket{g}_{N}=\sum_{\left\{ z_i \right\}}\Psi\para{ \left\{ z_i \right\} }\prod_{i=1}^{N}c_{z_{i}}^\dag \ket{0},\sd z=x+iy\cr
&&\Psi \para{ \left\{ z_i \right\} }=\braket{ \phi_{2m}^{1} \para{z_1}...\phi_{2m}^{1} \para{z_N} }={\rm Pf}\para{ \frac{1}{\para{z_i-z_j}^{m}}}.\cr
&&
\end{eqnarray}
The BCS-like wave function can be easily obtained by summing over a different number of electrons (with the same parity as that of $N$). The result would be:
\begin{eqnarray}
&&\ket{g}=\exp\para{ \frac{1}{\para{z_i-z_j}^{m}}c_{z_{i}}^\dag c_{z_{j}}^\dag} \ket{0}.
\end{eqnarray}

Now let us find the wave function in the presence of $2n$ vortices. Because the twist operator $\sigma_{1}\para{\eta}$ inserts a vortex at $\eta$, the wave function for $n$ vortices at $\eta_{j}$ where $i=\{1,..,2n\}$ and for a fixed number of electrons at $\left\{ z_i \right\}$ points where $i=\{1,..,N\}$, can be similarly written as:

\begin{eqnarray}
\Psi \para{\left\{ \eta_j\right\}; \left\{ z_i \right\} }=\braket{ \sigma_{1}\para{\eta_1}...\sigma_{1}\para{\eta_{n}}\phi_{2m}^{1} \para{z_1}...\phi_{2m}^{1} \para{z_N} }.&&\cr
&&
\end{eqnarray}
The above correlation function can be obtained by using the OPE between twist fields. As Eq. [\ref{OPE_10}] states, two twist fields can be fused in $2m$ different channels. In other words, although the positions of the vortices are known, the above correlation function can be evaluated in different ways. We can divide the $n$ vortices in $n/2$ pairs, fuse every two twist fields in a pair, and then find the correlation function. Again, we should be careful about the last pair, because after all fusions, we should end up with identity operator for a non-vanishing correlation function. Therefore, the last channel is dictated by the rest of the system. Therefore, there are $\para{2m}^{n/2-1}$ different conformal blocks, each corresponding to a wave function corresponding to the $\left\{q_{1},q_{2},...,q_{n/2-1}\right\}$ choice for fusion channels. Therefore, the  number of allowed degenerate ground states, i.e., the GSD, is $\para{2m}^{n/2-1}$ for $n$ vortices in the FTSC.   

\section{$\mathbb{Z}_{2m}$ gauge theory of the fractional topological superconductors at $1/m$ filling fraction}
Here, we analyze the gauge symmetry of the $\nu=1/m$ FTSCs and make a connection between non-Abelian FTSCs and the $Z_N$ rotor model discussed in Ref. \cite{Z_N_1}. In the FQH state, because of the number conservation, the gauge theory is described by a $U(1)$ gauge field. However, in the presence of pairing, this symmetry breaks down, and the number of electrons is conserved only modulus $2$. On the other hand, the electron operator can be imagined as the bound-state of $m$ anyon operators: $V_{e}=\phi_{2}^{m}$. Therefore, adding or removing two electrons is equivalent to adding or removing $2m$ anyons. In other words, the total number of anyons is conserved only modulus $2m$. Therefore, if we imagine anyons as the fundamental excitations of the ground state, the low-energy theory of the FTSC at $\nu=1/m$ filling fraction is invariant under the $Z_{2m}$ gauge transformation. On the other hand, as we discussed before, the quantum dimension of the topological defects (vortices) is $\sqrt{2m}$, which agrees with the prediction of $Z_{2m}$ rotor model. Using the latter formalism, You and Wen have shown the quantum dimension of the dislocations (vortices) in the $Z_{N}$ model to be $\sqrt{N}$.\cite{Z_N_1} From this consistency check, we speculate that $Z_{2m}$ symmetric FTSCs at $1/m$ filling fraction can be thought of as a realization of $Z_{2m}$ rotor model. 

\section{Summary, conclusion, and outlook}
In this paper, we introduced the notion of an FTSC. There are two classes of FTSCs, Abelian and non-Abelian. The Abelian FTSC is topologically the same as the Abelian FQH state, except that the electron number conservation is violated by injecting/inducing Cooper pairs into the FQH state. The superconducting order parameter is small in the Abelian FTSC so that we can smoothly evolve the Hamiltonian of an Abelian FQH into that of an FTSC without closing the energy gap. However, the non-Abelian FTSC is achieved upon closing and reopening the many-body energy gap of an Abelian FTSC by increasing the pairing amplitude. The resulting state allows non-Abelian FMFs with quantum dimensions equal to $\sqrt{2m}$, where $1/m$ refers to the filling fraction of the parent FQH state. We studied the braid statistics of these non-Abelions and derived the associated projective braid matrix. The wave function for the non-Abelian FTSCs was suggested through computing the conformal blocks of the edge CFT. We finally discussed the gauge theory of the effective low-energy Hamiltonian of the non-Abelian FTSC which happens to be a $Z_{2m}$ one. This observation makes a clear connection between our model of the non-Abelian FTSC with the $Z_{2m}$ rotor model with similar topological properties.

Due to the robustness of the zero modes in the topological phases against local perturbations of the Hamiltonian and disorder, the non-Abelian anyons are promising candidates for the topological quantum computation.  Information can be stored nonlocally in the degenerate ground states,  and braiding of quasiparticles causes unitary transformation on the ground state so can be used as the quantum gates.  However, it has been shown that when the square of the total quantum dimension of a non-Abelian system is integer, it cannot perform universal quantum computation by braiding only,\cite{Wang_2007_1} unless supplemented by measurement at the intermediate stages of the phase gate.\cite{Wang_2012_1,Barkeshli_2012_1} For non-Abelian FTSCs originated form Laughlin state at $\nu=1/m$, the square of the total quantum dimension, $D^2=\sum_{i=1}^{\rm N_{qp}}d_{i}^2$, where $N_{qp}$ is the number of quasiparticle, and $d_{i}$ their quantum dimension, is integer. Hence, they are capable of universal quantum computation with braiding and measurement. 

It is worth mentioning that the heat conductance in the FTSCs depends only on the difference between the central charge of the right- and left-mover boundary theories. Accordingly, we have $\kappa_{xy}=c\frac{\pi^2k_{\rm B}^2}{6\pi\hbar}T$. This equation can be proven by deriving the gravitational Chern Simons action of the FTSCs. 

Finally, we would like to comment on the topological field theory (TFT) of the bulk of a non-Abelian FTSC. Moore and Seiberg~\cite{M_S} have shown that for a system whose edge theory is given by $G/H$ orbifold CFT, where $H$ is a discrete subgroup of $G$, its bulk TFT is a Chern-Simons (CS) theory with $G \rtimes H$ gauge group. In our case, $G=U(1)$ and $H=Z_2$, so the bulk TFT of a non-Abelian FTSC at $\nu=1/m$ is described by an O(2) CS action at level $m$. Furthermore, the gauge symmetry of the $\nu=1/m$ FTSCs was discussed in this paper and shown to be a $Z_{2m}$ gauge symmetry. This observation makes our prediction about the quantum dimension of the FMFs more comprehensible through the results of the  Ref.~\cite{Z_N_1} for a $Z_{N}$ rotor model as well as those in Ref.~\cite{C_7}.

\noindent {\em Note added:} By the completion of this work, we became aware that the authors of Refs.~\cite{FTSC_1,FTSC_2,FTSC_3} have independently achieved similar results for the fractionalized Majorana edge states.

\section{Acknowledgement}
We gratefully acknowledge very useful discussions with Maissam Barkeshli, Eun-Ah Kim, and Xiao-Gang Wen.


\end{document}